\documentclass[12pt,titlepage,a4paper]{article}
\usepackage[dvips]{graphicx}
\usepackage{epsfig}
\usepackage{amsfonts}
\usepackage{amsmath}

\begin{document}
\title{Acoustic wave propagation in two-phase heterogeneous porous media}
\author{
J.I. Osypik\thanks{email jam@srcc.msu.ru}, N.I. Pushkina\thanks
{email N.Pushkina@mererand.com}, Ya.M. Zhileikin\thanks{email jam@srcc.msu.ru}\\
M.V.Lomonosov Moscow State University, Research Computing Center, \\ Vorobyovy Gory, 
Moscow 119991, Russia} 
\date{ }
\maketitle
\begin{abstract}
The propagation of an  acoustic wave through two-phase 
porous media with spatial variation in porosity is studied. The evolutionary 
wave equation is derived, and the propagation of an acoustic wave 
is numerically analyzed in application to marine sediments with
different physical parameters.  
\end{abstract}
\section*{Introduction}

Investigation of acoustic-wave propagation in two-phase porous media, 
marine sediments in particular, finds an increasing interest in studying 
physical properties of such media. There has been a significant amount of 
publications on the propagation of sound in the sea floor. Acoustic-wave 
propagation in sediments is controlled by intrinsic properties of a
sediment, characterized by a number of physical parameters. One of the 
parameters that significantly influences sound propagation in a sediment 
is the porosity. The porosity indicates relative amounts of solid 
and liquid fractions in a sediment and hence it determines 
the frame bulk and shear moduli and through this the acoustic-wave speed. 
The variations in the medium properties can arise due to random packing 
of inhomogeneous sediment grains. 
In Refs. \cite{Hef1, Hef2, Hef3, Hef4} dispersion and 
acoustic-wave scattering from randomly varying heterogeneities in 
the poroelastic medium properties such as the porosity or the 
frame bulk modulus has been studied experimentally and theoretically. 
In Ref. \cite{Hef4} perturbation theory is used to derive a poroelastic 
wave equation which describes the first-order scattering by 
the heterogeneities of a medium. The scattering of energy from 
heterogeneities accounts for additional losses of a sound wave 
propagating through a poroelastic medium. It was shown that random 
variations in the parameters have more significant effects on the
sound propagation through a consolidated medium than through 
a sand sediment.

In the present paper, we study the influence of porosity variations 
in space on the propagation of a  plain acoustic wave in 
marine sediments. We shall describe the porosity of a sediment 
and similarly some other physical parameters as fluctuations 
about their average values. If the fluctuations are small 
the sound field is assumed to change slowly at the 
wave-length scale, and we shall use the  method 
of slowly varying form of the wave (see Ref. \cite{zz}) 
to develop a poroelastic evolutionary wave equation 
in a heterogeneous medium. This method is similar to the method  
of a slowly varying amplitude widely used in studying nonlinear 
wave interactions in media with dispersion, for instance, in 
studying light-wave interactions in nonlinear optics. To analyze 
different cases of spatial variation in porosity computer simulation 
of the obtained evolutionary equation is performed.

\section[]{Derivation of approximate wave equations \\ 
for finite-amplitude acoustic waves \\ 
in marine sediments}

To derive the approximate evolution wave 
equation we start from the continuity equations for the densities and 
momenta of the liquid and solid phases of a sediment composed of a rigid 
frame and pores filled with water, see Ref. \cite{BN}. These equations 
are equivalent in the main features to the equations developed by Biot 
\cite {Bi1, Bi2, Bi}, 
but they are presented in a somewhat different form, we shall not list 
them here. On the basis of these equations, in Ref. \cite{Zh} the 
equations for the densities of the liquid and solid phases, 
$\rho_f$ and $\rho_s$, were derived (in this paper we don't take 
diffraction and nonlinearity into account):

\begin{eqnarray}
\left(1-\frac{m}{\rho_fc^2G}\right)\frac{\partial\rho_f}
{\partial\tau}-
\frac{\nu}{\rho_sc^2G}\frac{\partial\rho_s}{\partial\tau}=
\nonumber\\-c\left(1+\frac{m}
{\rho_fc^2G}\right)\frac{\partial\rho_f}{\partial x}-
\frac{\nu}{\rho_scG}\frac{\partial\rho_s}{\partial x}; \label{eq:fl}
\end{eqnarray}
\begin{eqnarray}
\left(1-m-\frac{k+4/3\mu+\nu^2/G}{\rho_sc^2}\right)
\frac{\partial\rho_s}{\partial\tau}-\frac{m\nu}
{\rho_fc^2G}\frac{\partial\rho_f}{\partial\tau}=\nonumber\\
-c\left(1-m-\frac{k+4/3\mu+\nu^2/G}{\rho_sc^2}\right)
\frac{\partial\rho_s}{\partial x}-\frac{m\nu}
{\rho_fcG}\frac{\partial\rho_f}{\partial x}. \label{eq:sol}
\end{eqnarray}
In these equations the following variables are used,  
\begin{equation}
x'=\epsilon x \label{eq:x}
\end{equation}
and the moving coordinate 
\begin{equation}
\tau=t-x/c. \label{eq:t}
\end{equation}
(In Eqs. (\ref{eq:fl}), (\ref{eq:sol}) and everywhere below the 
primes for $x$ are omitted.)
In the relation (\ref{eq:x}) the small parameter $\epsilon$ is 
introduced as
\begin{equation}
\epsilon \sim v_x/c \sim u_x/c \sim \delta \rho_f/\rho_f
\sim \delta \rho_s/\rho_s,
\end{equation}
here $c$ is the speed of sound in the sediment and ${v}$, ${u}$ 
are the hydrodynamic velocities of the liquid and solid phases, 
$\delta \rho_f$, $\delta \rho_s$ are the deviations from equilibrium 
values of the densities of the liquid and solid phases.
In Eqs. (\ref{eq:fl}), (\ref{eq:sol}) the left-hand sides  
are of the order of $\sim\epsilon$, the right-hand side terms are 
of the order of $\sim\epsilon^2$. 
The introduction of the new variables (\ref{eq:x}), (\ref{eq:t}) 
actually signifies the application of the method of slowly
varying wave profile.\\
In the equations (\ref{eq:fl}), (\ref{eq:sol}) we wrote $\rho_n, \rho_s$
instead of $\delta\rho_n, \delta\rho_s$, $m$ is the porosity; 
\[G=\frac{1-m}{k_s}+\frac{m}{k_f}-\frac{k}{k^{2}_{s}},\] 
where  $k_f$, $k_s$ and $k$ are the bulk moduli of the fluid, 
mineral grains constituting the frame, and of the frame itself;
$\mu$ is the shear modulus of the frame; $\nu=1-m-k/k_s$. 
 
Let us now describe the porosity of a sediment as fluctuations 
about an average value, $m=m_0+\triangle m$. Alongside 
the porosity the bulk and shear moduli of the frame and the 
speed of sound should similarly fluctuate about their average values,
$k=k_0-\triangle k,\,\,\,\,\,\mu=\mu_0-\triangle\mu,\,\,\,\,\,
c=c_0-\triangle c$. The deviations of the moduli and speed of sound 
from their equilibrium values have an opposite sign as to the 
fluctuations of the porosity since with the increase of the 
porosity  the sediment frame becomes softer.

Let us eliminate one of the variables, $\delta \rho_f$ or 
$\delta \rho_s$, from the left-hand sides of Eqs. (\ref{eq:fl}), 
(\ref{eq:sol})), 
(let it be, e.g., $\delta \rho_s$) by subtracting one equation 
from the other one. In the right-hand sides the quantity 
$\delta \rho_s$ is expressed through $\delta \rho_f$ with the 
formula which is valid to an accuracy $\sim\epsilon$,
\begin{equation}
\delta\rho_s =\left(\frac{\nu}{\rho_sc^2G}\right)^{-1}
\left(1-\frac{m}
{\rho_fc^2G}\right)
\delta\rho_f. \label{eq:lin}
\end{equation}

Note, that Eqs. (\ref{eq:fl}), (\ref{eq:sol})  
allow two independent longitudinal 
modes, the so called fast and slow waves. As it is shown in Ref. 
\cite{St}, the slow wave (unlike the fast one) is a strongly 
attenuated diffusion mode, and it does not contribute significantly 
to the sound field. In this approximation we arrive at the 
equation for an acoustic wave in a sediment with 
parameters, that vary with distance:
\begin{eqnarray}
\left\{2(1-m)\left[1-\left(\frac{c_f}{c}\right)^2\right]-
\frac{\nu^2}{m}
\frac{\rho_f}{\rho_s}\left(\frac{c_f}{c}\right)^2+\left
(1-m-\frac{k+4/3\mu}
{\rho_sc^2}\right)
\left[1+\left(\frac{c_f}{c}\right)^2\right]\right\}\frac
{\partial
\rho_f}{\partial x}+\nonumber\\
\left\{m+\frac{k+4/3\mu}{\rho_sc^2}\left(\frac{c_f}{c}\right)^2
\left[2\frac{\rho_f}{\rho_s}\nu(1-2m)+2-3m-3\frac{k+4/3\mu}{\rho_sc^2}
\right]\right\}\delta(x)\frac{\partial\rho_f}{\partial \tau}+\nonumber\\
a_1D_\tau\rho_f
 =0.\label{eq:D}
\end{eqnarray}
To obtain these equations we took into account that in 
sand sediments the bulk modulus $k_s$ of quartz grains is much 
greater than that of the pore water, and in this case $G$ can be 
evaluated as $G \approx m/k_f$, provided $m$ is not close to zero.  
In Eq. (\ref{eq:D}) we have used the quantity $\delta(x)$ 
which absolute value would be around 0.1,
\begin{equation}
\delta(x)\sim\frac{\triangle m}{m_0}\sim\frac{\triangle k}{k_0}
\sim\frac{\triangle\mu}{\mu_0}\sim\frac{\triangle c}{c_0}.
\end{equation}
In Eq. (\ref{eq:D}) the term $a_1D_\tau\rho_f$ that accounts for 
dissipation is introduced. $D_\tau$ is the dissipation linear operator 
in the variable $\tau$ which is characterized by the property
\begin{equation}
D_\tau e^{i\omega\tau}=\alpha(\omega)e^{i\omega\tau},\label{op} 
\end{equation}
where $\alpha$ is real and positive and it has the meaning 
of an amplitude attenuation coefficient if the coefficient 
$a_1$ is taken equal to the coefficient at $\partial\rho_f/\partial x$.
The relation (\ref{op}) defines the action of this operator on any 
function of the variable $\tau$ which can be represented by a 
Fourier series or integral. An algebraic expression for $\alpha(\omega)$ 
is a combination of physical parameters (complex bulk and shear frame moduli 
included) of a sediment, and it includes 
the frequency correction function introduced by Biot \cite {Bi}. 

In Introduction it was noted that acoustic wave scattering from 
randomly varying heterogeneities 
in the poroelastic medium properties manifests itself in the increase 
of the sound field attenuation. In this paper, we shall not consider 
sound scattering from heterogeneities, including it implicitly 
in the dissipation term. 

\section[]{Computer simulation of acoustic wave\\ propagation
in heterogeneous absorbing\\ marine sediments} 

To study numerically the propagation of acoustic waves in  absorbing
marine sediments with  parameters varying with distance we consider 
Eq. (\ref{eq:D}) presented in a concise form:
\begin{equation}
a_1\frac{\partial\rho_f}{\partial x}+a_2\frac{\partial\rho_f}{\partial \tau}
\delta(x)+a_1D_\tau\rho_f=0.  \label{C}
\end{equation}



As it is seen from Eq. (\ref{eq:D}) the coefficients $a_1,a_2$ are the algebraic combinations of 
physical parameters of a sediment. It is convenient to divide Eq. (\ref{C}) by $a_1$,

\begin{equation}
\frac{\partial\rho}{\partial x}+\frac{a_2}{a_1}\delta(x)\frac{\partial\rho}{\partial \tau}
+D_\tau\rho=0.  \label{11}
\end{equation}

Let the density boundary value be
$$
\rho|_{x=0} = A\rho_0,\qquad  A=10^{-3}\mbox{--}10^{-5}.
$$

Introducing a new variable  $\theta = 10^{4}\tau $ 
we have $\rho (2\pi 10^4\tau) = \rho (2\pi \theta )$.
 
Solving Eq. (\ref{11}) we are to find the function $\rho(2\pi \theta)$ 
periodic in the variable $\theta $ with the period $1$. The boundary condition is taken 
to be a harmonic function  
$$
\rho _0 = -\sin (2\pi \theta). 
$$

In Eq. (\ref {11}) it is convenient to normalize the functions and the variables
except the variable $x$ measured in centimeters. We obtain the  equation

\begin{equation}
\frac{\partial\rho}{\partial x}+C\delta'(x)\frac{\partial\rho}{\partial \theta}
+D_\theta\rho=0.  \label{12}
\end{equation}
\begin{equation} \label{13}
\rho |_{x=0} = -\sin(2\pi \theta),   
\end{equation}
where 

$$
C=\frac{\varepsilon}{a_1}\frac {b}{c}10^4\delta_0, \qquad \varepsilon = \pm 1, 
\qquad \delta (x) = \delta_0\delta'(x), 
$$
$$
\delta '\varepsilon (x)\in [0,1], \qquad \delta _0\in [0.1,0.2], 
$$

$$
D_\theta = 10^4D_\tau      
$$

Eq. (\ref{12}) describes in fact two processes, the change of the wave phase and the wave dissipation. 
To solve it the so called splitting method \cite{Godunov} is applied.  

Consider a simple example,

\begin{equation}
\frac{du}{dx} = Au + Bu, \qquad u|_{x=0} = u_0      \label{14}
\end{equation}
and calculate $u(h)$, were $h$ is the step in $x$.

We can divide the problem into two parts
$$
\frac{dv}{dx} = Av,\qquad   v|_{x=0} = u_0,   
$$
$$
\frac{dw}{dx} = Bw, \qquad  w|_{x=0} = v(h).
$$
For smooth solutions the equality $w(h) = u(h) + O(h^2)$ holds true. So,
we can obtain the solution to Eq. (\ref{14}) solving two more simple problems. 

The solution of the equation 
\begin{equation}  \label{15}
\frac{\partial \rho}{\partial x} + C\delta' (x)\frac {\partial \rho}{\partial \theta} = 0,
\end{equation}
$$
\rho |_{x=0} = \rho_0,
$$
satisfying periodic boundary condition can be received with the difference
schemes of the "angle"  type. The stencil of this difference scheme is 
defined by the characteristic equation
$\displaystyle{\frac{d\theta }{dx} = C\delta' (x)}$. Since 
$\delta' (x)$ is positive  or equal to zero the direction of the characteristics
depends on the sign of the coefficient $C$. If $C\ge 0$ the "right angle" difference scheme
 is stable, if  $C\le 0$ the "left angle" difference scheme is
also stable. The condition binding the steps in $x$ and $\theta$
is of  the form $\displaystyle{h\le \frac{z}{|C|}}$ ($h$ and $z$ are the steps for
 $x$ and $\theta $ axes respectively).

If $\rho |_{x=0}$ is represented  as the Fourier series $\rho |_{x=0} = \sum \nu_m(0)e^{2\pi im\theta}$,
  Eq.  (\ref{15}) can be solved in an explicit form: 
$$
\rho(x,\theta) = \sum\nu_m(x)e^{2\pi im\theta}
$$ 
with
$$
\nu_m(x) = \nu_m(0)e^{-2\pi imC\int_0^x\delta' (\xi)d\xi}.
$$
Since $m = \pm 1$ we have  $\rho (x,\theta) = -\sin (2\pi (\theta -\mu))$, $\mu = 
C\displaystyle{\int _0^x\delta' (\xi)d\xi}.$
This means that the solution of (\ref{15}) gives a shift of the phase equal to $2\pi \mu $. 
The phase shift moves to the right if   $C$ is positive and to the left if $C$ is negative. 

Let us consider the equation
$$
\frac{\partial \rho}{\partial x} + D_\tau \rho = 0.
$$
$D_\tau$ is the linear dissipation operator :
$$ 
D_\tau  \Longrightarrow  D_\theta  = 10^4\left |\frac {\partial \rho}{\partial \theta }\right |\alpha' ,
$$
$10^4\alpha'= \alpha$ is the attenuation coefficient. 

This relation defines 
the action of this operator on a function of the variable $\theta$ represented by the Fourier series 
$\rho = \sum \nu _me^{2\pi im\theta}$.
As a result we obtain the equation
$$
\frac {d\nu _m}{dx} = -\frac{\alpha'}{2\pi}10^4|d_m|, \qquad \quad  d_m \approx 2\pi m,
$$
from which one gets
$$
\nu _m = e^{-\frac{\alpha'}{2\pi}10^4|d_m|x}.
$$
Since $m = \pm 1$, we have 
$$
\rho(x,\theta) = -\sin (2\pi \theta)e^{-\alpha' 10^4x}.
$$

The parameter  $\displaystyle{L_d = \frac{1}{\alpha' 10^4}}$ is the propagation  
distance. 


Consider some examples describing the transformation of harmonic acoustic 
waves propagating along $x$. It will be seen that the change of the porosity 
with distance leads to the phase shift of the initial acoustic wave.

In Figures  1 and 2 the graphs of the functions  $\rho |_{x=0}$ and \\
$\delta' (x) = 0.667 + 0.333\sin(10^{-1}\pi x)$ are presented.

We take  $C = 0.652\cdot 10^{-2}$ and
$10^4\alpha' = \alpha \simeq 0.45\cdot 10^{-2}cm^{-1}$  
(see experimental data in Refs. \cite{chot, stoll, buch}).

The number of nodes of the variable  $\theta$ in the interval  $[0,1]$ is $64$. 
This number 
of nodes is sufficient for  approximating a harmonic function.
The step of the spatial variable $x$ is $0.5$, that corresponds to the magnitude 
of  $|C|$ and the  stability condition.

The graphs of $\rho (x_i,\theta)$ at $x_i = 25,50,75,100$  
are presented in Figs 3--6. A positive value of $C$ gives the   
wave phase shift to the  right. If $C$ is negative (with the same module), the wave phase 
 shifts to the left.




 In the above examples the  porosity has an oscillating character 
  (Fig.~2) and satisfies the inequality    
 $0.331\le \delta'(x)\le 1$. If $C$ is positive the curve 
 shifts to the right more quickly for larger 
$\delta'(x)$ values and slowly for smaller $\delta'(x)$ values. 
Negative $C$ values lead to shifting the initial 
curve   to the left in a similar way. The porosity as an 
oscillating function has been chosen as an  example.
 
 Now we shall consider more realistic cases of an irregular spatial variation 
 in a sediment porosity. Take for example  arbitrary continuous $\delta'(x)$ functions 
 of the form presented in  Figs. 7--9.

Fig. 7: let $C>0$. In Figs. 7.1--7.6 the function $\rho(x_i,\theta)$ 
is presented  for six $x_i$ values: $x_i$= 30, 40, 50, 60, 70 and 80.
 If  $0\le x_i\le 40$ the initial curve $\rho(x_i,\theta)$ does not shift,
that is its phase does not change; if $40<x_i<60$ it moves  to the right 
with an increasing speed;
if $60\le x_i\le 100$ it moves with a constant high speed.   

Fig.~8. The grafs for the initial function $\rho(x_i,\theta)$ are not listed here, 
since they are in a sence similar to those in Figs. 7.1--7.6. 
At $0\le x_i\le 40$ the phase changes so that the function $\rho(x_i,\theta)$ shifts  
to the right with a constant speed. 
At  $x_1>40$ the speed decreases and at $x_i\ge 60$ it goes to zero. 

Fig.~9: This case 
qualitatively repeats the cases of Figs. 
7--8 where the curves go
respectively up or down.
The speed of the initial curve shift slows down at $0\le x_i < 50$ 
and at  $x_i > 50$ the shift moves with an accelerating speed.  

In conclusion 
consider the case of a random pore size distribution (Fig. 10) that can arise,
 for instance, due to random packing of the sediment  nonuniform 
 grains. Such porosity distribution can be presented as a random digital array,
 and we shall interpolate it with a continuous function, see Fig. 11. 
This continuous function in its turn can be  approximately  considered as a series of curves 
of the types  presented in Figs. 7--8.  That is, the consideration given above 
for Figs. 7--8 can be 
applied to  each section of this function.

\section[]{Conclusion} 

Evolutionary wave equation to describe acoustic wave propagation in a two-phase porous 
media with spatial variations in porosty is derived. Computer simulation of the obtained 
equation is performed to analyze diverse cases of the porosity variations that lead  
to phase shifts in the initial harmonic acoustic wave.

\pagebreak

\centerline{FIGURES}

\bigskip

\begin{figure}[ht]
\includegraphics[width=0.35\linewidth]{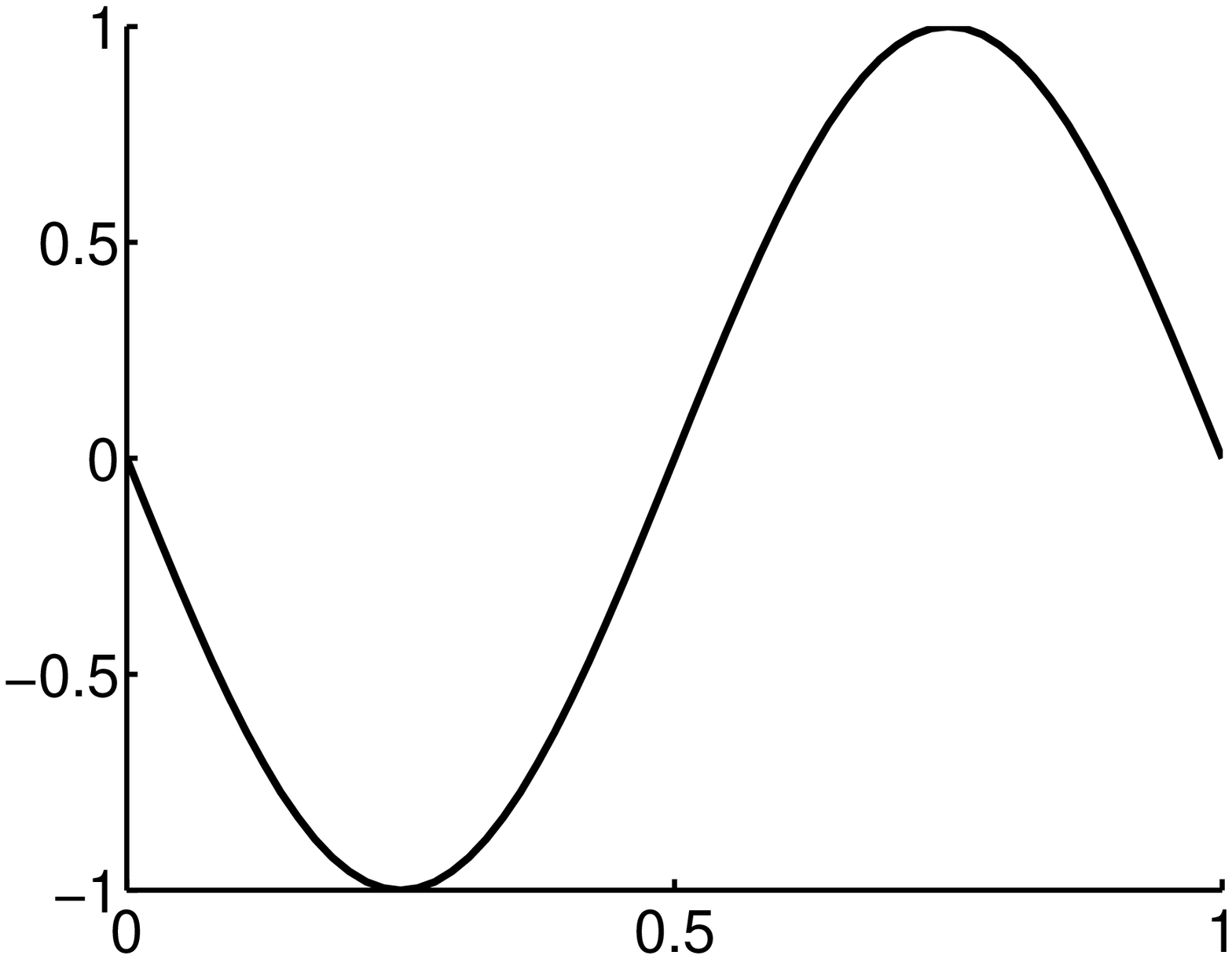}
\hspace{10mm}
\includegraphics[width=0.35\linewidth]{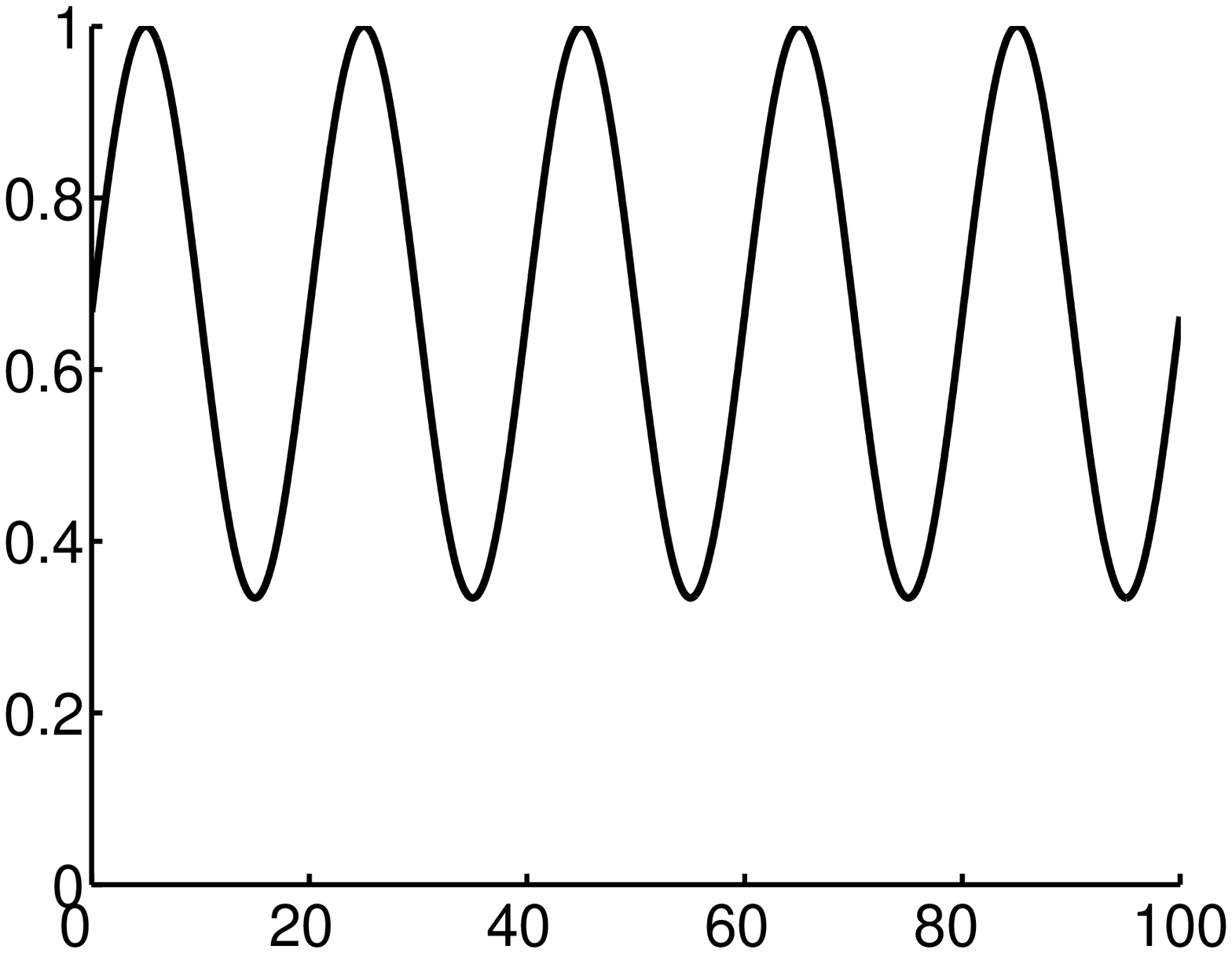}

{Fig.~1. The function $\rho (\theta )|_{x=0}$}  \hspace{20mm} {Fig.~2. The function $\delta ' (x)$}

\end{figure}

\begin{figure}[h]

\includegraphics[width=0.40\linewidth]{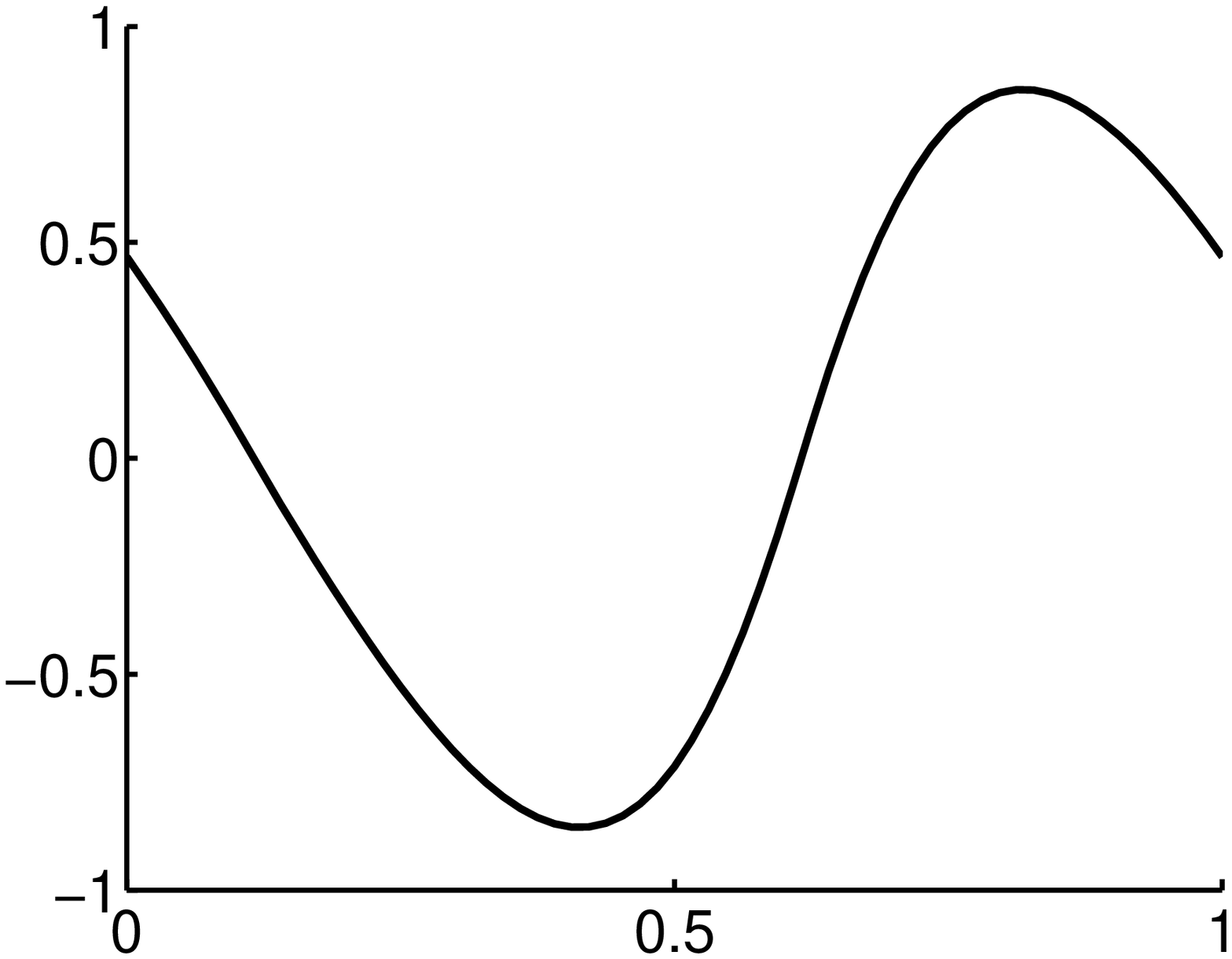}
\hspace{10mm}
\includegraphics[width=0.40\linewidth]{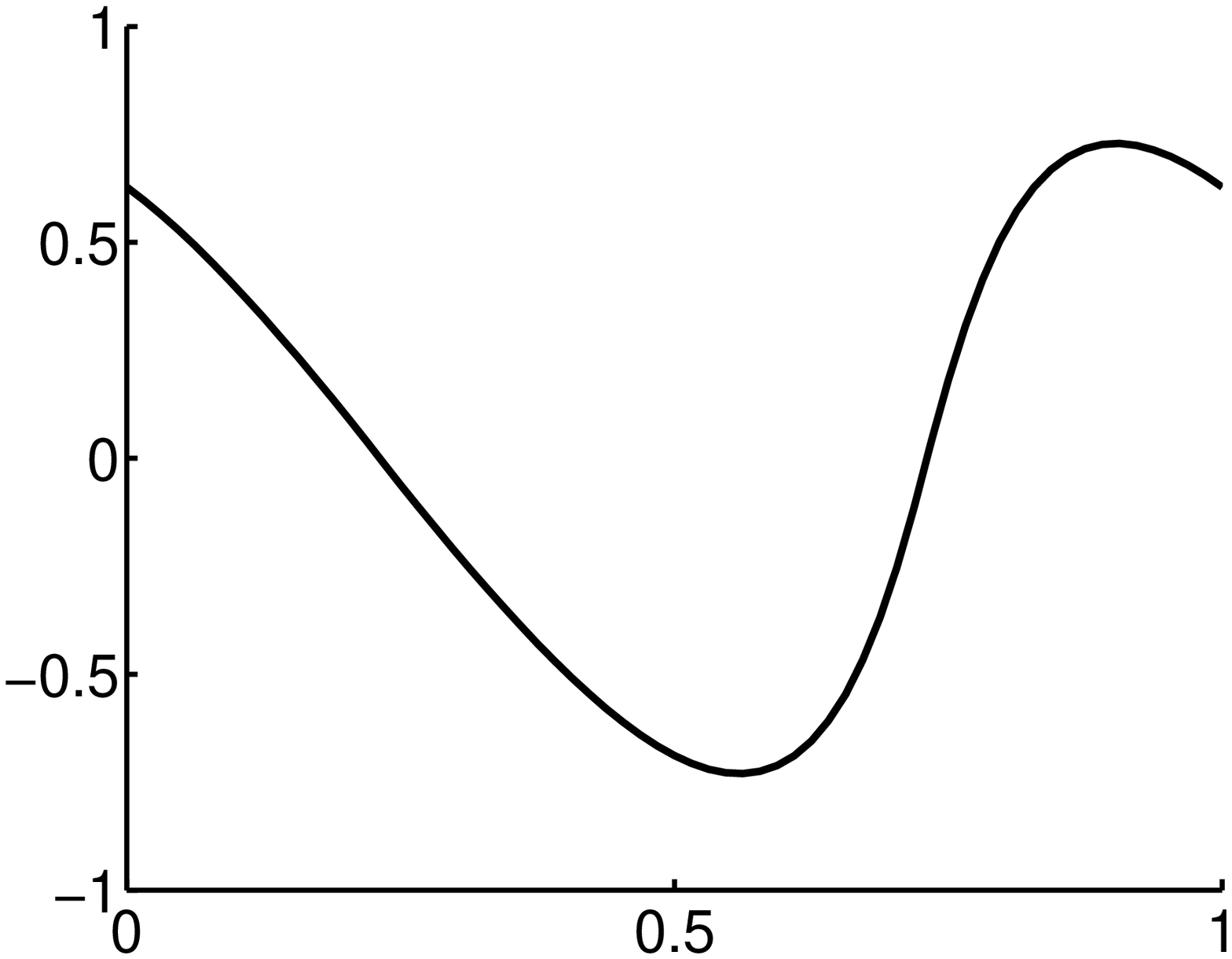}

{Fig.3. The function $\rho(x_i,\theta)$ at $x_i = 25$}\hspace{5mm}
{Fig.4. The function $\rho(x_i,\theta)$ at $x_i = 50$}

\includegraphics[width=0.40\linewidth]{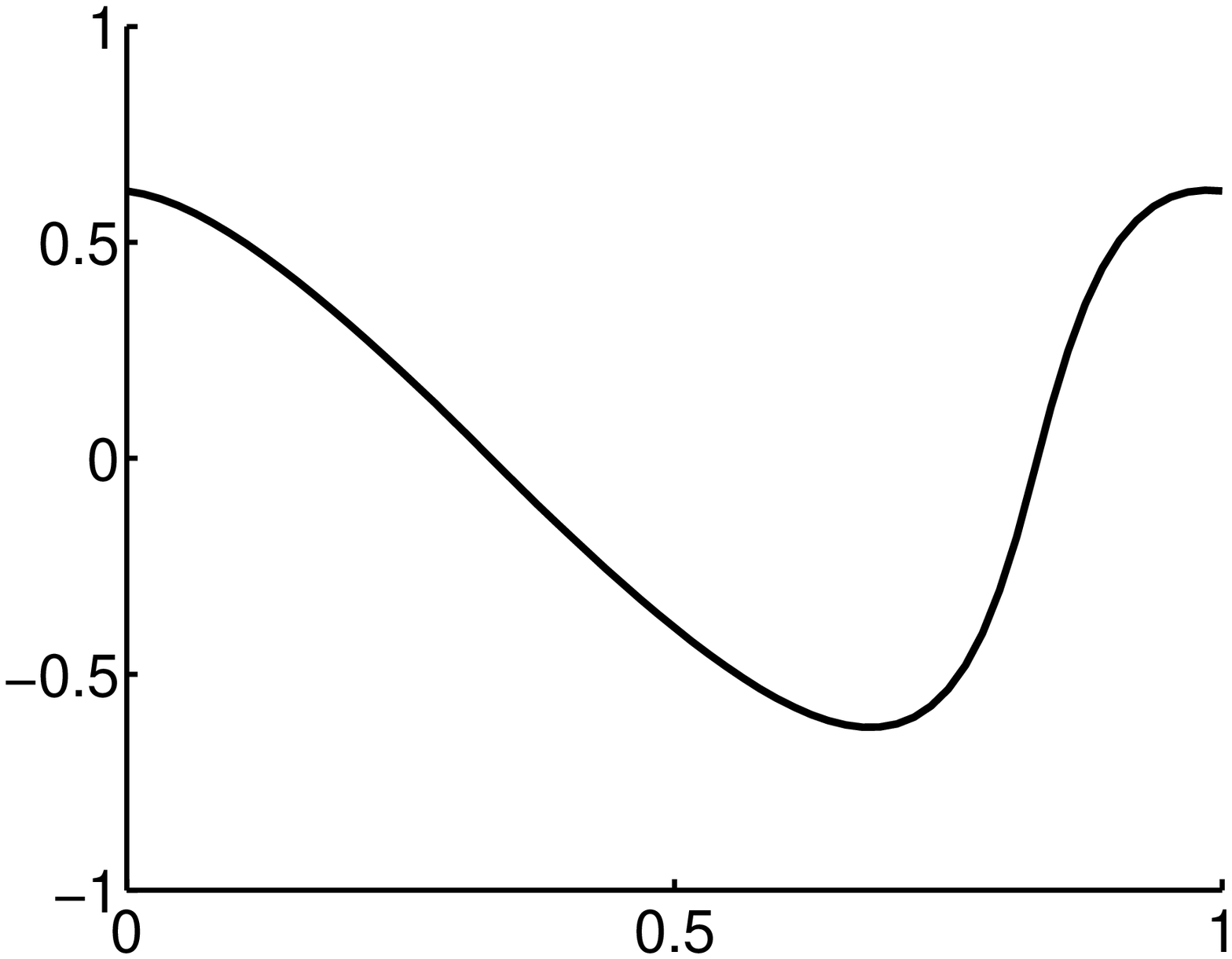}
\hspace{10mm}
\includegraphics[width=0.40\linewidth]{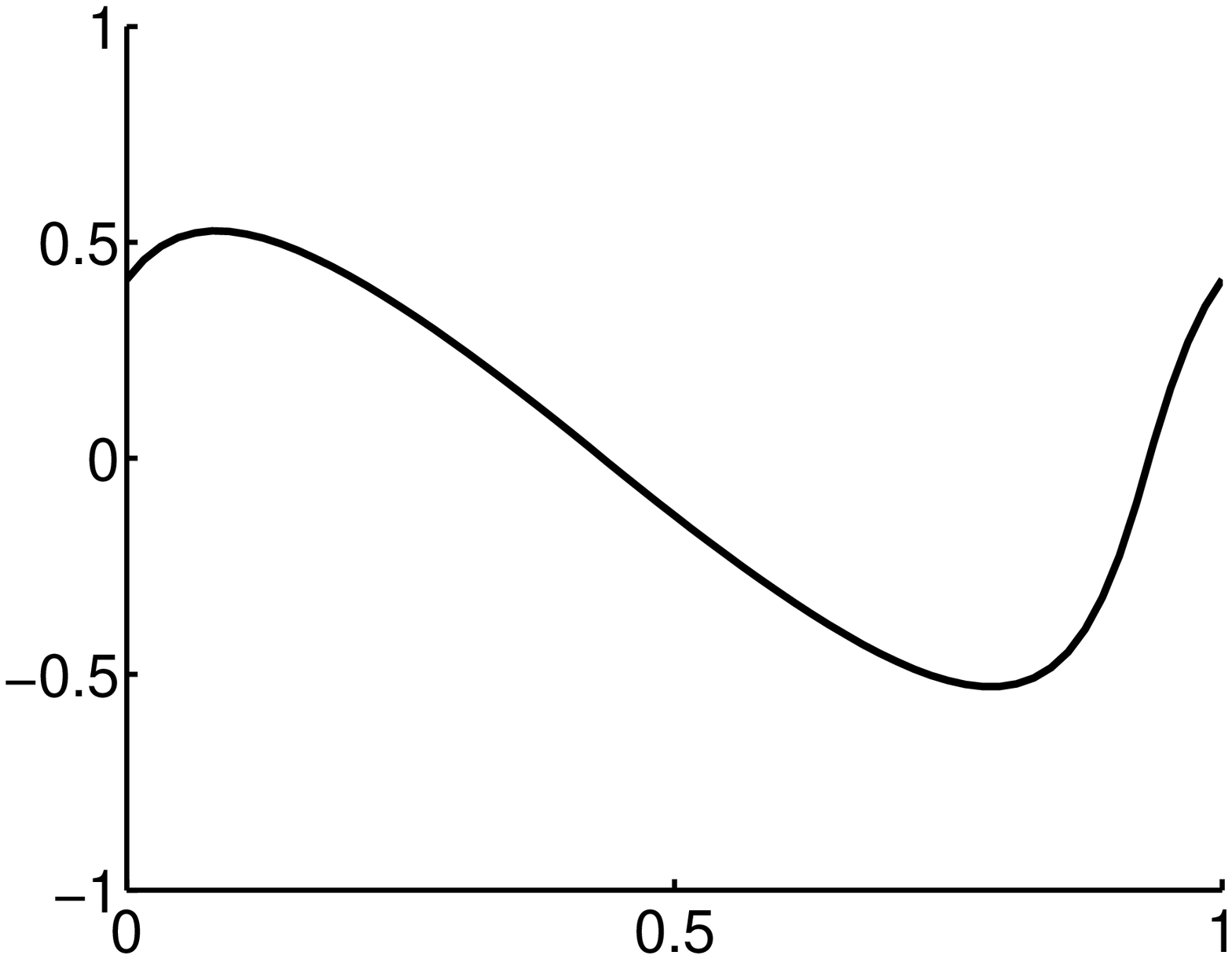}

{Fig.5. The function $\rho(x_i,\theta)$ at $x_i = 75$}\hspace{5mm}
{Fig.6. The function $\rho(x_i,\theta)$ at $x_i = 100$}

\end{figure}

\begin{figure}[!h]
\vspace{10mm}

\includegraphics[width=0.4\linewidth]{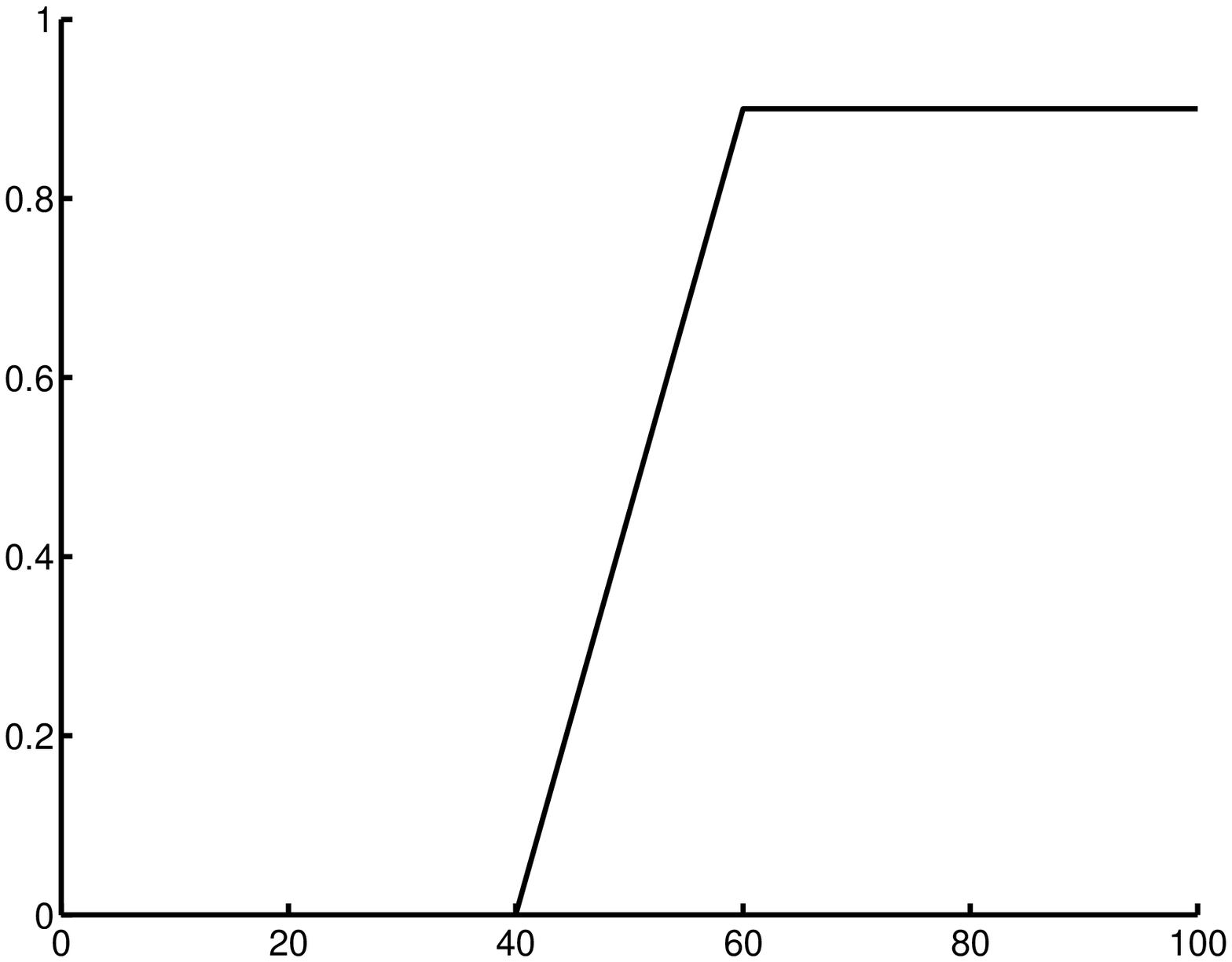}
\hspace{10mm}
\includegraphics[width=0.4\linewidth]{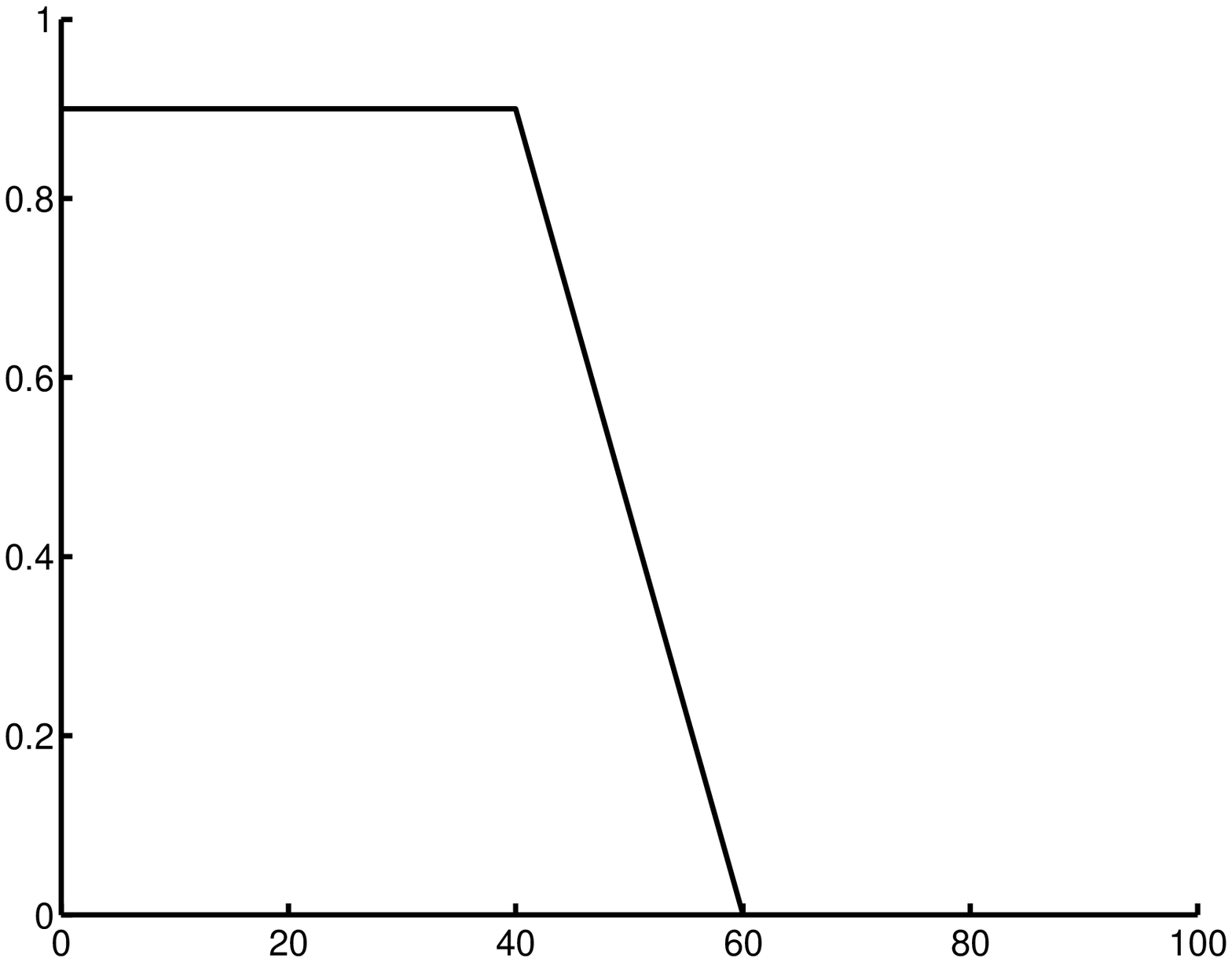}

\hspace{25mm}  {Fig.~7}  \hspace{60mm} {Fig.~8}

\vspace{5mm}

\includegraphics[width=0.4\linewidth]{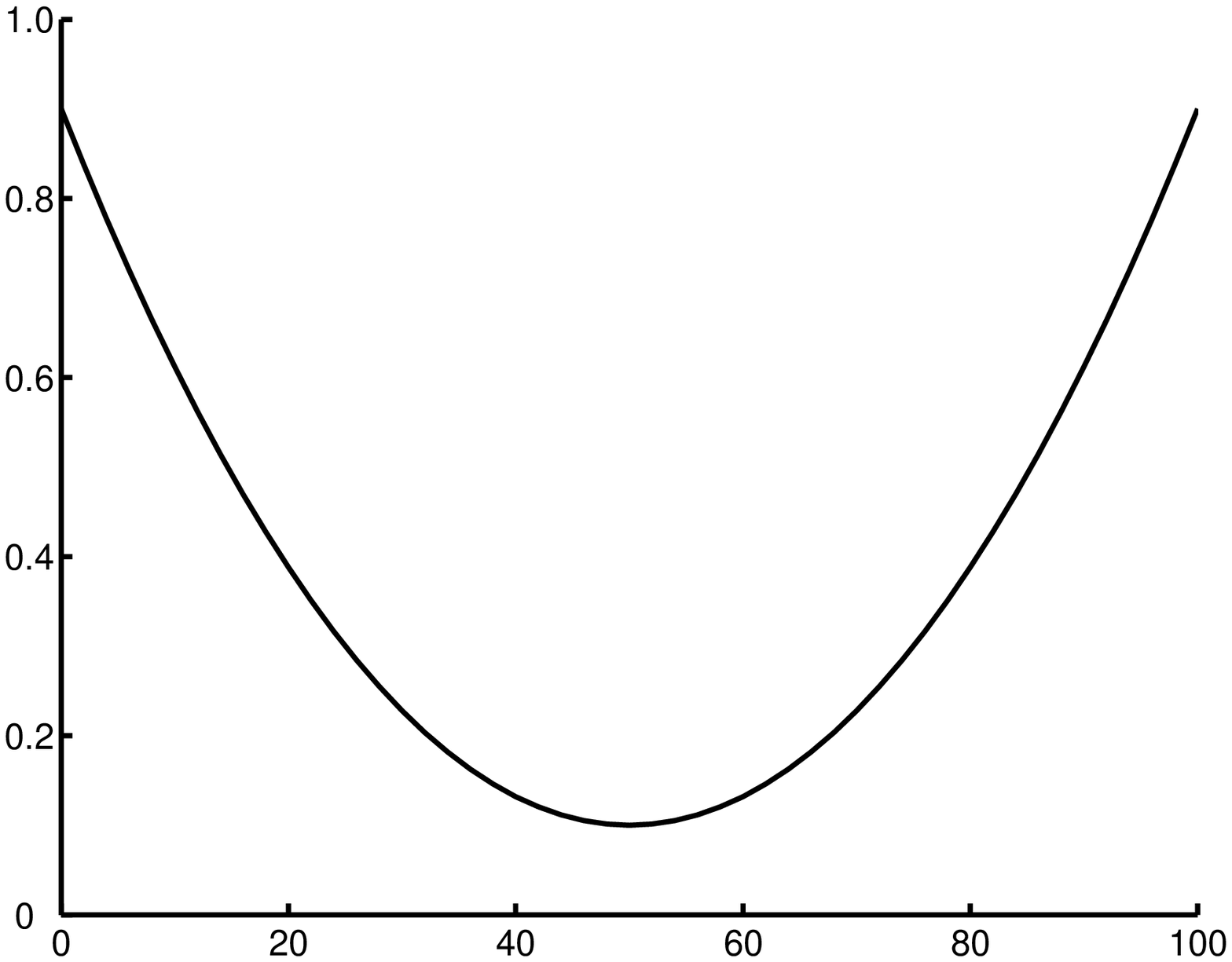}
\hspace{10mm}
\includegraphics[width=0.4\linewidth]{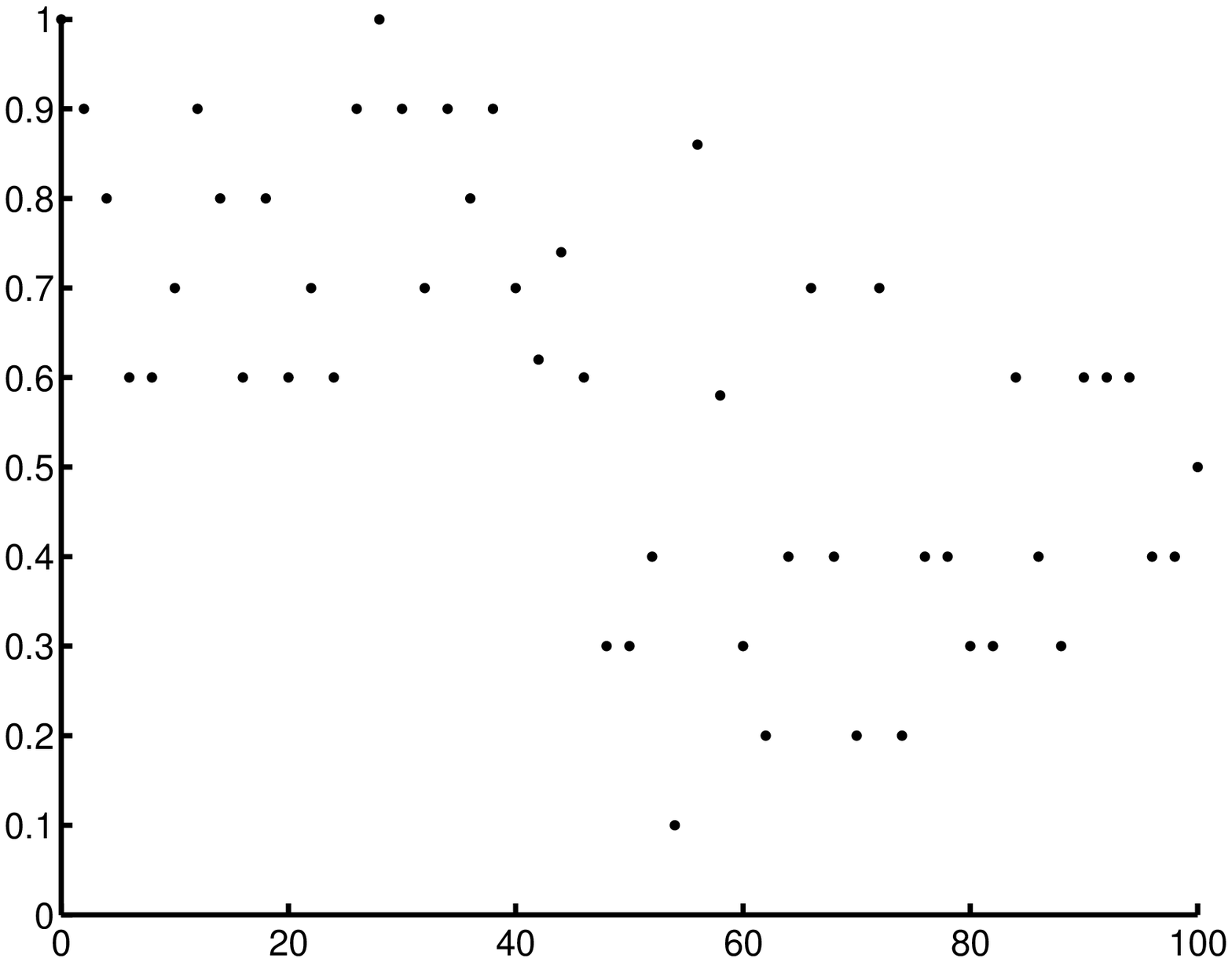}

\hspace{25mm} {Fig.~9}  \hspace{60mm} {Fig.~10}

\end{figure}

\begin{figure}[!h]
\vspace{10mm} 
\hspace{50mm}
\includegraphics[width=0.4\linewidth]{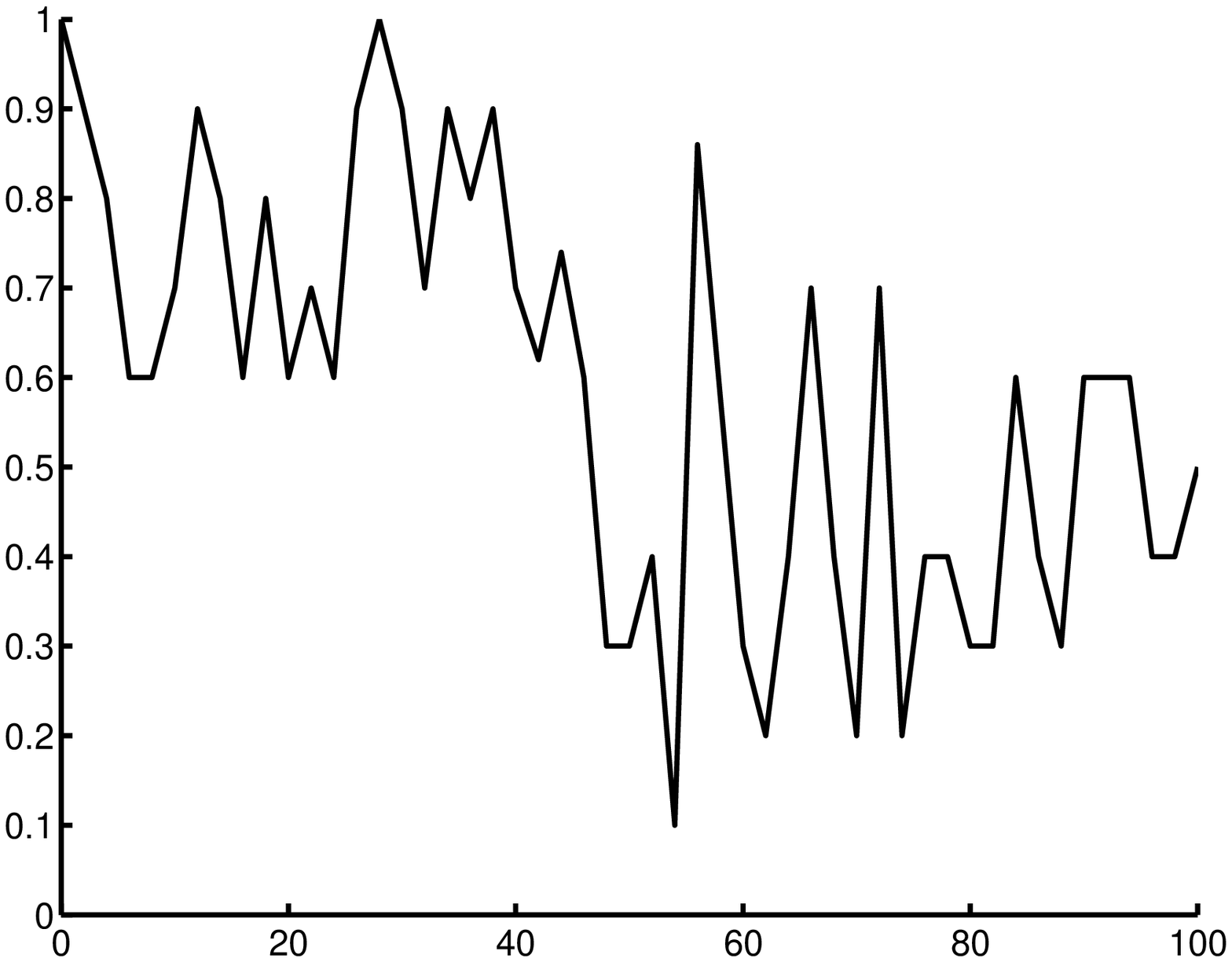}

\hspace{75mm}  {Fig.~11}  
                                       
\end{figure}

\begin{figure}[!h]

\includegraphics[width=0.4\linewidth]{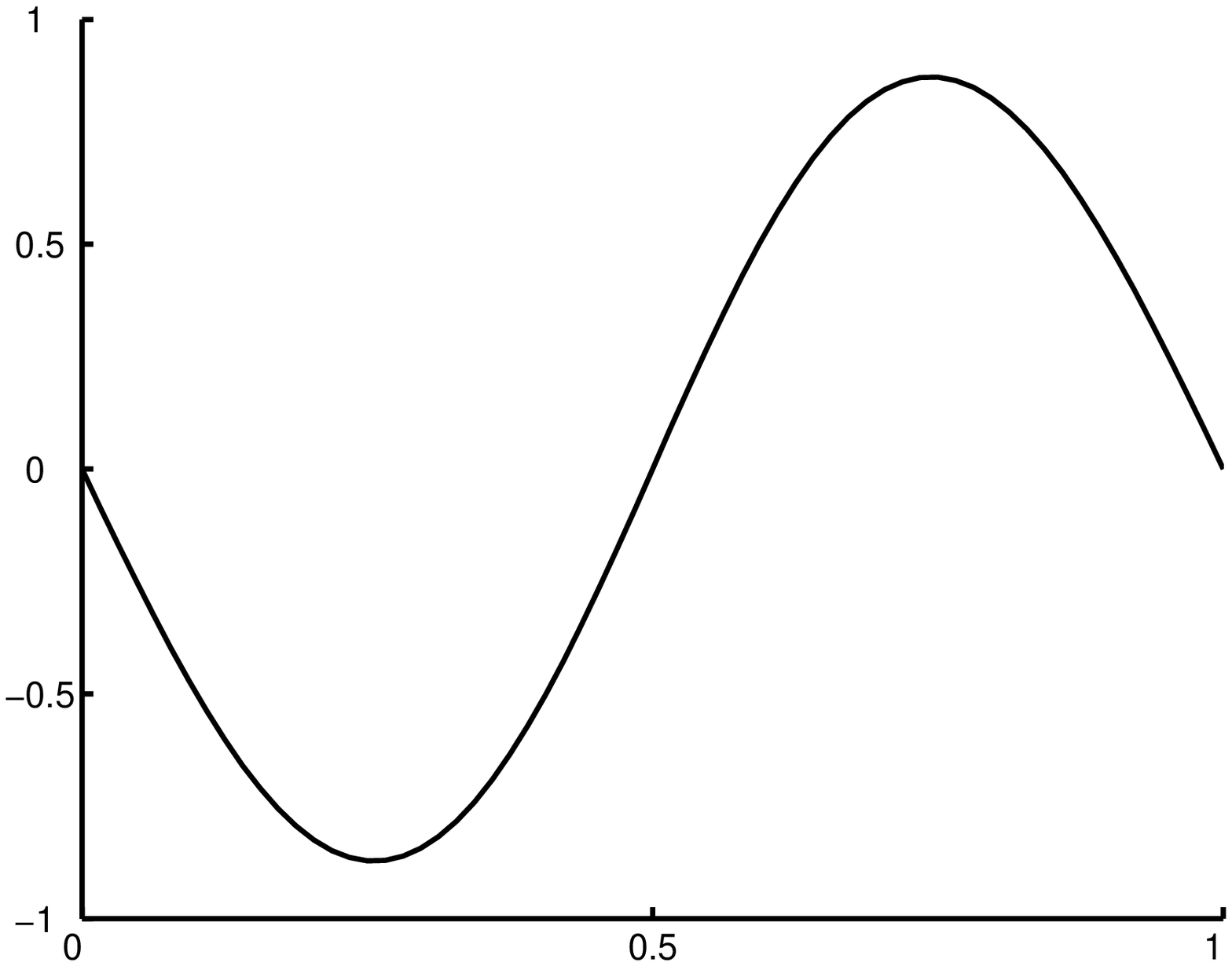}
\hspace{10mm}
\includegraphics[width=0.4\linewidth]{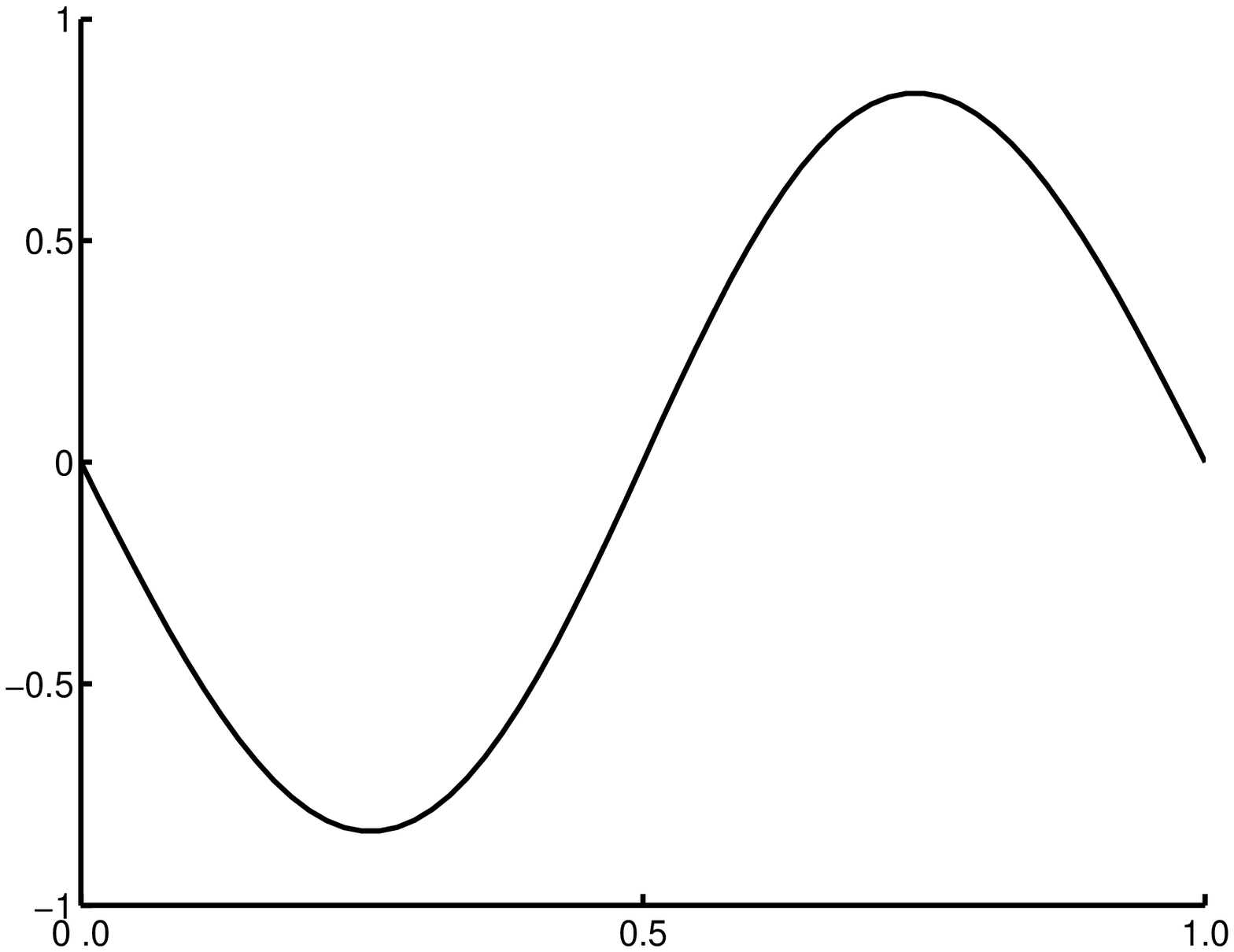}

{Fig.~7.1. The function $\rho (x_i,\theta)$ at $x_i = 30$}
\hspace{5mm} {Fig.~7.2. The function $\rho (x_i,\theta)$ at $x_i = 40$}

\vspace{5mm}

\includegraphics[width=0.4\linewidth]{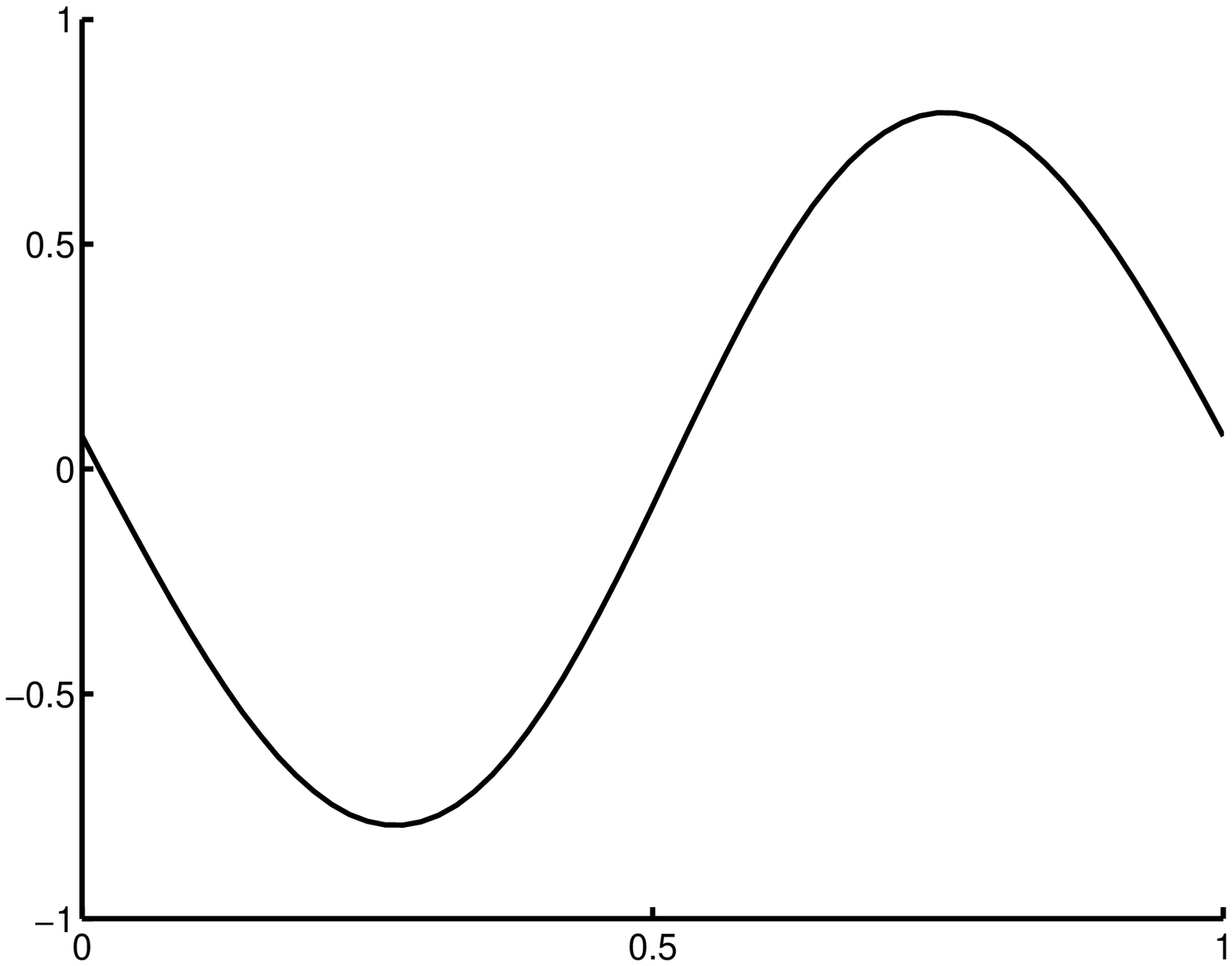}
\hspace{10mm}
\includegraphics[width=0.4\linewidth]{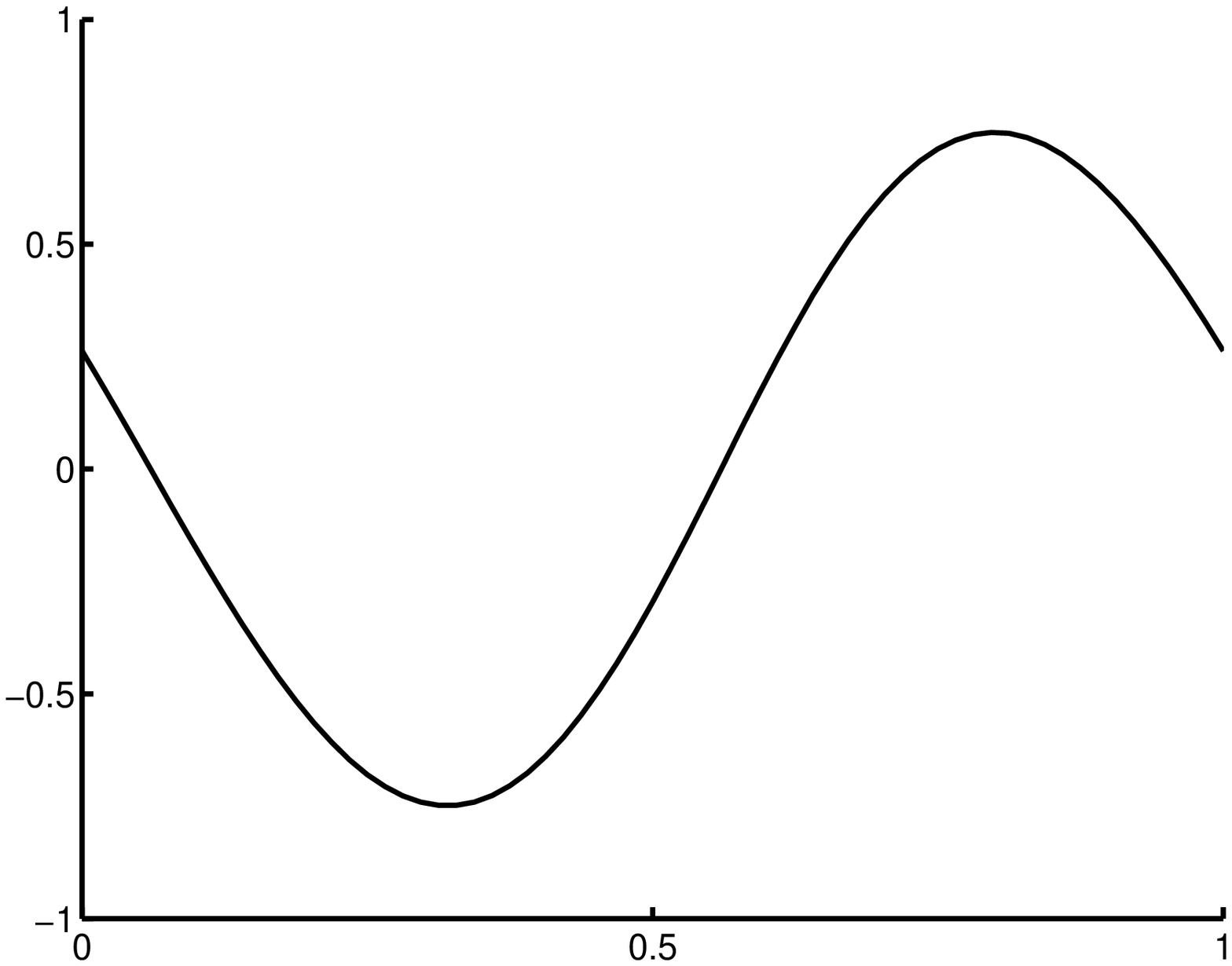}

{Fig.~7.3. The function $\rho (x_i,\theta)$ at $x_i = 50$}
\hspace{5mm}{Fig.~7.4. The function $\rho (x_i,\theta)$ at $x_i = 60$}

\includegraphics[width=0.4\linewidth]{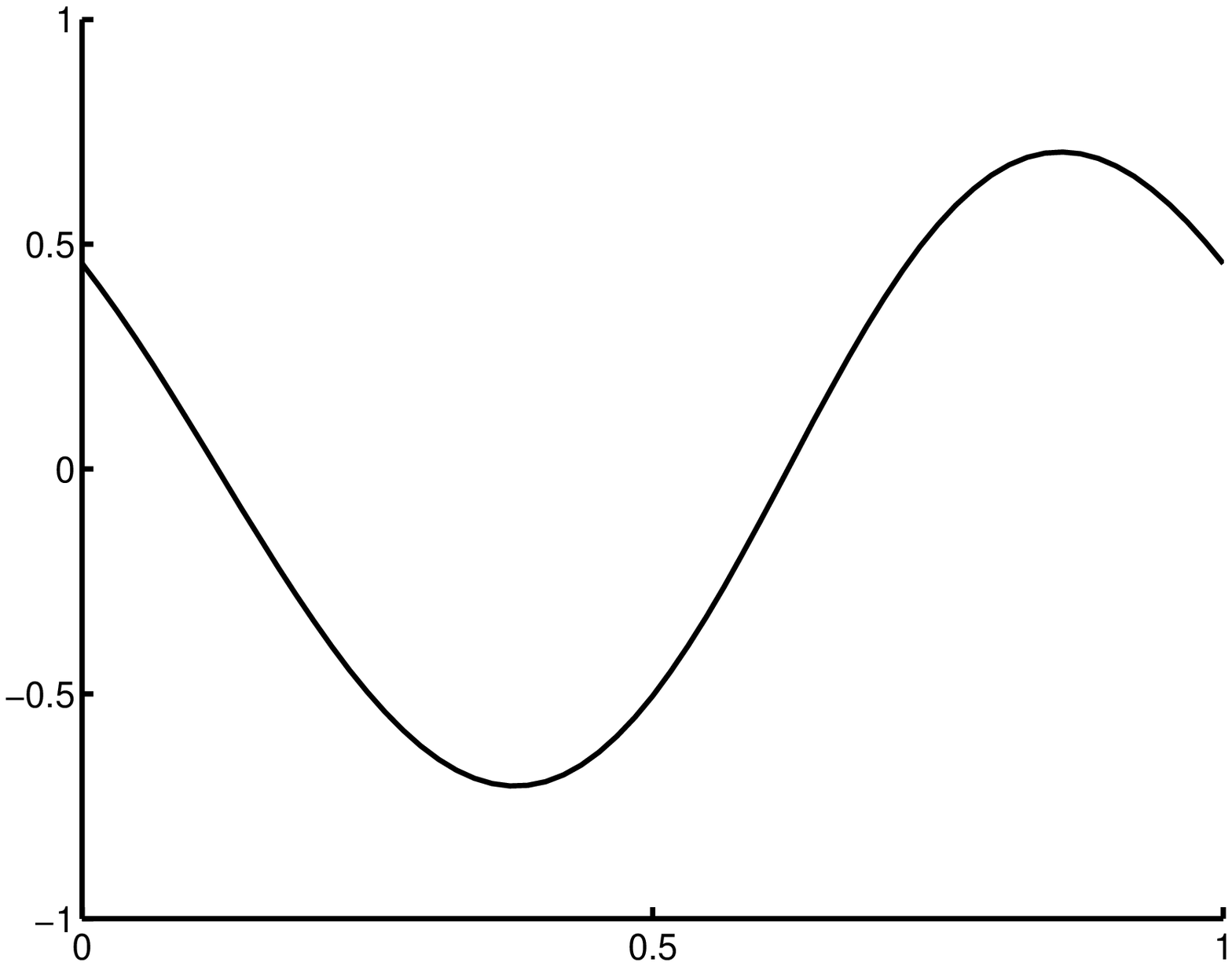}
\hspace{10mm}
\includegraphics[width=0.4\linewidth]{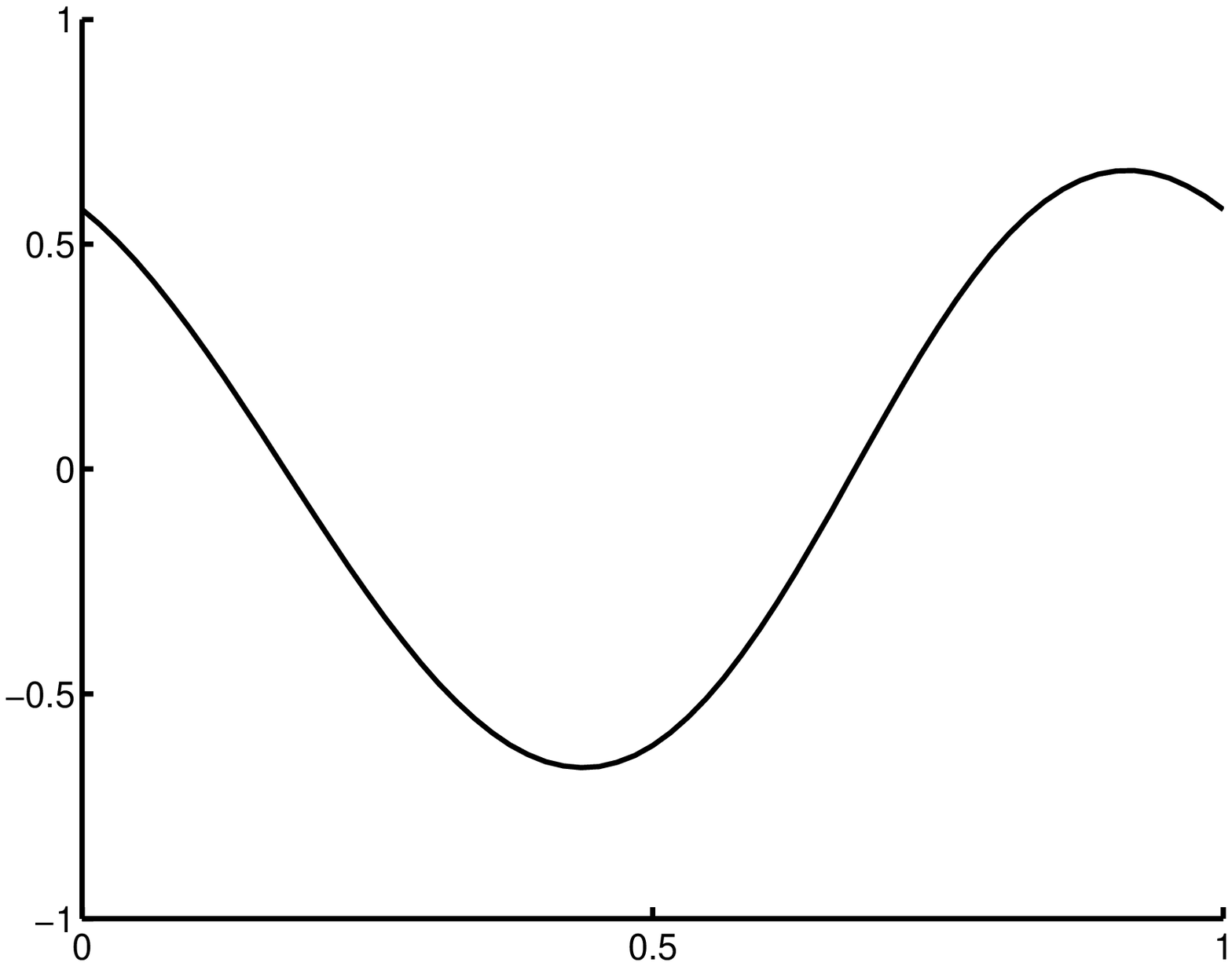}

{Fig.~7.5. The function $\rho (x_i,\theta)$ at $x_i = 70$}
\hspace{5mm}{Fig.~7.6. The function $\rho (x_i,\theta)$ at $x_i = 80$}

\end{figure}

\end{document}